\documentclass[twocolumn,aps,prl,superscriptaddress,floatfix]{revtex4-1}

\usepackage{xcolor}
\usepackage{graphicx}
\usepackage{dcolumn}
\usepackage{bm}
\usepackage{amsmath}
\usepackage{upgreek}
\usepackage{braket}
\usepackage{comment}

\usepackage[T1]{fontenc}
\usepackage[unicode=true, pdfusetitle, bookmarks=true, bookmarksnumbered=false, bookmarksopen=false, breaklinks=false, pdfborder={0 0 1}, backref=false, colorlinks=false]
{hyperref}
\hypersetup{
colorlinks, linkcolor=blue, citecolor=blue, urlcolor=blue
}

\newcommand{\angstrom}{\mbox{\normalfont\AA} }

\newcommand{\ham}{\hat{H}}
\newcommand{\F}{\hat{F}}

\newcommand{\Neq}{\hat{N}_\x{eq}}
\newcommand{\dt}{\Delta t}
\newcommand{\eps}{\varepsilon}
\newcommand{\x}[1]{\mathrm{#1}}
\newcommand{\trace}[1]{\mathrm{Tr}\{ #1 \}}
\newcommand{\tee}{\tau_\x{ee}}

\DeclareMathOperator{\sech}{sech}

\newcommand{\tep}{\tilde{\varepsilon}}

\newcommand{\tH}{\widetilde{H}}

\begin{document}

\preprint{APS/123-QED}

\title{Real-Time Out-of-Equilibrium Quantum Dynamics in Disordered Materials}

\author{Luis M. Canonico}
\email{luis.canonico@icn2.cat}
\affiliation{Catalan Institute of Nanoscience and Nanotechnology (ICN2), CSIC and BIST, Campus UAB, Bellaterra, 08193 Barcelona, Spain}
\author{Stephan Roche}
\affiliation{Catalan Institute of Nanoscience and Nanotechnology (ICN2), CSIC and BIST, Campus UAB, Bellaterra, 08193 Barcelona, Spain}
\affiliation{ICREA, Instituci\'o Catalana de Recerca i Estudis Avançats, 08070 Barcelona, Spain}
\author{Aron W. Cummings}
\email{aron.cummings@icn2.cat}
\affiliation{Catalan Institute of Nanoscience and Nanotechnology (ICN2), CSIC and BIST, Campus UAB, Bellaterra, 08193 Barcelona, Spain}

\date{\today}

\begin{abstract}
\noindent
We report a linear-scaling numerical method for exploring nonequilibrium electron dynamics in systems of arbitrary complexity.
Based on the Chebyshev expansion of the time evolution of the single-particle density matrix, the method gives access to nonperturbative excitation and relaxation phenomena in models of disordered materials with sizes on the experimental scale.
After validating the method by applying it to saturable optical absorption in clean graphene, we uncover that disorder can enhance absorption in graphene and that the interplay between light, anisotropy, and disorder in nanoporous graphene might be appealing for sensing applications. Beyond the optical properties of graphene-like materials, the method can be applied to a wide range of large-area materials and systems with arbitrary descriptions of defects and disorder.
\end{abstract}


\maketitle
Understanding nonequilibrium charge dynamics is crucial for assessing electronic, energy, and angular momentum transport in quantum materials \cite{ferrari2015science_etal}. For instance, charge carriers propagating in graphene under optical excitation are quickly driven to a highly nonequilibrium state \cite{song2015energy, nair2008fine, OpticalCondGraphene, UniversalOpticalCondGraphite, dawlaty2008measurement, tielrooij2013photoexcitation, electroniccoolingGraphene, EnergyRelaxation, massicotte2021hot}, and these “hot carriers” serve as the functioning principle for ultra-compact sensing and communication devices \cite{bonaccorso2010graphene, romagnoli2018graphene, OptimizingPTE}. Examples of such devices are sensitive THz antennas \cite{castilla2019fast}, and saturable absorbers usable in mode-locked lasers \cite{bao2009atomic, bao2011monolayer, baek2012efficient}.
Moreover, recent developments allow for the bottom-up growth of atomically-precise nanomaterials such as nanoporous graphene (NPG) \cite{moreno2018bottom}, an array of covalently-linked graphene nanoribbons that exhibits a sizable band gap \cite{moreno2018bottom, calogero2018electron} and strong in-plane transport anisotropy \cite{calogero2018electron, calogero2019quantum, mortazavi2019nanoporous, alcon2024tailoring}. Such bottom-up engineering can thus imbue 2D carbon materials with properties appealing for sensing applications present in other 2D materials such as transition metal dichalcogenides \cite{wang2012electronics} or black phosphorous \cite{xia2014rediscovering}.

The study of out-of-equilibrium processes requires a nonperturbative description of the carrier dynamics for both the excitation of the system and the treatment of disorder. Disorder is a fundamental part of device fabrication and can drastically alter the energy transfer dynamics and optical response of any material, including graphene \cite{graham2013photocurrent, alencar2014defect, halbertal2017imaging}, NPG \cite{alcon2024tailoring}, transition metal dichalcogenides (TMDs) \cite{dai2019enhanced,liang2021defect,bianchi2024engineering} and graphene/TMD heterobilayers \cite{hotger2023photovoltage}. Semiclassical \cite{xing2010physics, song2015energy, electroniccoolingGraphene} and microscopic \cite{malic2011microscopic, winzer2012absorption, marques2004time, yamada2018time} methods have been developed to study nonequilibrium processes. Furthermore, real-time extensions of TD-DFT theory are capable of describing the excited carrier dynamics of solids under various excitations and capture dynamics beyond Ehrenfest approximations \cite{wang2020natural} by including a Boltzmann weight on the transitions and restoring detailed balance. In particular, these approaches have been used in the study of hot-carrier cooling after photoexcitation \cite{liu2021critical,wang2025effect}. Still, despite their success, these techniques are not currently suited to deal with a nonperturbative description of disorder on typical experimental length scales. Meanwhile, real-space linear-scaling methods can study electron transport in systems containing hundreds of millions of atoms, reaching experimental scales, while treating disorder nonperturbatively \cite{Fan2021linear, kite}. To date, these methods have been limited to closed quantum systems in which energy is conserved, or to the linear-response regime where the external interactions cannot change the electron density, making them unable to track nonequilibrium quantum dynamics. Thus, the theoretical investigation of nonequilibrium phenomena in realistic systems calls for approaches that incorporate the best of both worlds.

In this Letter, we present a numerical method for simulating out-of-equilibrium electron dynamics in systems comprised of millions of atoms while treating disorder nonperturbatively.
To illustrate the method's capabilities, we use it to study saturable absorption in graphene and NPG in the presence of electrostatic disorder. We demonstrate that it can describe optical absorption nonperturbatively by tracking the time- and energy-resolved carrier distribution. Our results show excellent agreement with previous experiments in clean graphene, and indicate that electrostatic disorder may improve the performance of graphene saturable absorbers. We also demonstrate that NPG exhibits optical anisotropy that can be enhanced by weak disorder. While we have focused on optical absorption, this method is completely generalizable, making it suitable for studying far-from-equilibrium electron dynamics in various materials and systems approaching experimental length scales while considering nonperturbative disorder effects.

\textit{Methodology} -- The core aspect of the methodology is the time evolution of the density matrix. To explore nonequilibrium dynamics, we want to track its evolution under a time-varying Hamiltonian.
We start with the ground state density matrix, represented by the Fermi operator $\F(\ham_0,T_0,\mu_0) = [ 1 + \exp((\ham_0-\mu_0)/k_\x{B}T_0) ]^{-1}$, where $\ham_0$ is the Hamiltonian in equilibrium while $T_0$ and $\mu_0$ are the initial electronic temperature and chemical potential \cite{baer1997sparsity, baer1997chebyshev, ordejon1998order, goedecker1999linear}. Under a Hamiltonian that evolves from its equilibrium state $\ham_{0}\rightarrow\ham(t)$, we then calculate the time-dependent electron occupation as
\begin{equation}
\braket{n(\eps,t)} = \frac{\trace{\hat{U}^\dagger(t,0) \delta(\ham(t)-\eps) \hat{U}(t,0) \F(\ham_0,T_0,\mu_0)}}{\trace{\delta(\ham(t)-\eps)}},
\label{eq:occupation}
\end{equation}
where $\hat{U}(t_1,t_0) = \hat{\cal T} \exp\left\{ -\frac{\x{i}}{\hbar} \int_{t_0}^{t_1} \ham(t') dt' \right\}$ is the time-ordered evolution operator and $\delta(...)$ is the Dirac delta operator, which projects the occupation onto energy $\eps$.

We approximate the trace as an average over random phase states \cite{WeisseKPM, Fan2021linear}, $
\trace{\hat{A}} \approx \frac{1}{R} \sum_{r=0}^R \braket{\psi_r | \hat{A} | \psi_r},$ where in the site basis $\ket{\psi_r} = \frac{1}{\sqrt{N}} \left[ \x{e}^{\x{i}\xi_1} \, ... \, \x{e}^{\x{i}\xi_N} \right]^T$, $\xi_n$ is a random number evenly distributed in $[0,2\pi)$ and $N$ is the number of sites. This random phase approximation introduces a numerical noise that scales as $\delta\hat{A} \propto 1 / \sqrt{RN}$, and in general, for a good approximation one only needs a few states, $R \ll N$ \cite{WeisseKPM, Fan2021linear}. In all of our calculations, we used $R=32$ random phase states.

In Eq.\ \eqref{eq:occupation} we have to determine the time evolution of the vectors $\ket{\psi_r(t)} \equiv \hat{U}(t,0) \ket{\psi_r}$ and $\ket{\psi_{r}^{\F}(t)} \equiv \hat{U}(t,0) \F(\ham_0,T_0,\mu_0) \ket{\psi_r}$. For short simulation time steps $\dt$, $\ket{\psi_r(t)}$ can be evolved using the instantaneous Hamiltonian, such that $\ket{\psi_{r}(t+\dt)} = \hat{U}(t+\dt,t)\ket{\psi_{r}(t)} \approx \exp(-\x{i} \ham(t) \dt / \hbar) \ket{\psi_r(t)}$. In contrast, $\ket{\psi_{r}^{\F}(t)}$ contains the full history of the electron occupation. In addition to the accumulation of energy driven by $\ham(t)$, we would like to track other processes, not necessarily captured by $\ham(t)$, that modify the electron occupation. For example, high-energy electrons may emit phonons, dumping energy to the lattice and relaxing to a thermal distribution at the lattice temperature \cite{electroniccoolingGraphene, graham2013photocurrent, pogna2021hot_etal}. Meanwhile, carrier-carrier scattering in graphene conserves the total energy of the electrons but drives them toward a thermal distribution that may be much higher than the lattice temperature \cite{brida2013ultrafast, song2015energy, massicotte2021hot}.

To that end, we propose a modified time evolution for $\ket{\psi_{r}^{\F}(t)}$ that captures both a time-varying Hamiltonian and relaxation toward a given carrier distribution,
\begin{align}
\frac{\partial}{\partial t} \ket{\psi_{r}^{\F}(t)} = &-\frac{\x{i}}{\hbar} \ham(t) \ket{\psi_{r}^{\F}(t)} \nonumber \\
&- \frac{1}{\tau} \left( \ket{\psi_{r}^{\F}(t)} - \Neq(t)\ket{\psi_{r}(t)} \right),
\label{eq:evolution}
\end{align}
where $\Neq(t)$ is the instantaneous equilibrium distribution toward which the electrons relax, over a time scale $\tau$. As stated above, this could be driven by electron-phonon or electron-electron scattering, or could be any other distribution governed by the physics of the system.

In general, $\Neq(t)$ can vary in time and may depend on the instantaneous carrier distribution, as in the case of electron-electron scattering. Additionally, complex terms in the Schr\"odinger equation lead to vanishing carrier density \cite{midgley2000complex}. We thus perform the time evolution of Eq.\ \eqref{eq:evolution} in two steps. First we compute $\ket{\psi_{r}(t+\dt)} = \hat{U}(t+\dt,t) \ket{\psi_r(t)}$ and $\ket{\psi_{r}^{\F}(t+\dt)} = \hat{U}(t+\dt,t) \ket{\psi_{r}^{\F}(t)}$. Next, from these updated vectors we determine a new carrier distribution $\Neq(t + \dt)$ and mix it with $\ket{\psi_r^{\F}(t + \dt)}$ using forward time stepping. Equation \eqref{eq:evolution} then becomes
\begin{align}
\ket{\psi_r^{\F}(t+\dt)} &\approx \left( 1-\frac{\dt}{\tau} \right) \hat{U}(t+\dt,t)\ket{\psi_r^{\F}(t)} \nonumber \\
&+ \frac{\dt}{\tau} \Neq(t+\dt) \hat{U}(t+\dt,t) \ket{\psi_r(t)}.
\label{eq:evolution2}
\end{align}

In the application to graphene below, we assume $\Neq$ is given by electron-electron scattering, which drives $\braket{n(\eps,t)}$ toward a thermal distribution over a time scale $\tee$.
Thus, $\Neq(t) = \F(\ham(t), T(t), \mu(t))$, where at each time step we determine the temperature $T(t)$ and the chemical potential $\mu(t)$ to ensure the conservation of energy and carrier density \cite{SupplementaryMaterial}.

\begin{figure}[th]
\centering
\includegraphics[width=0.975\linewidth]{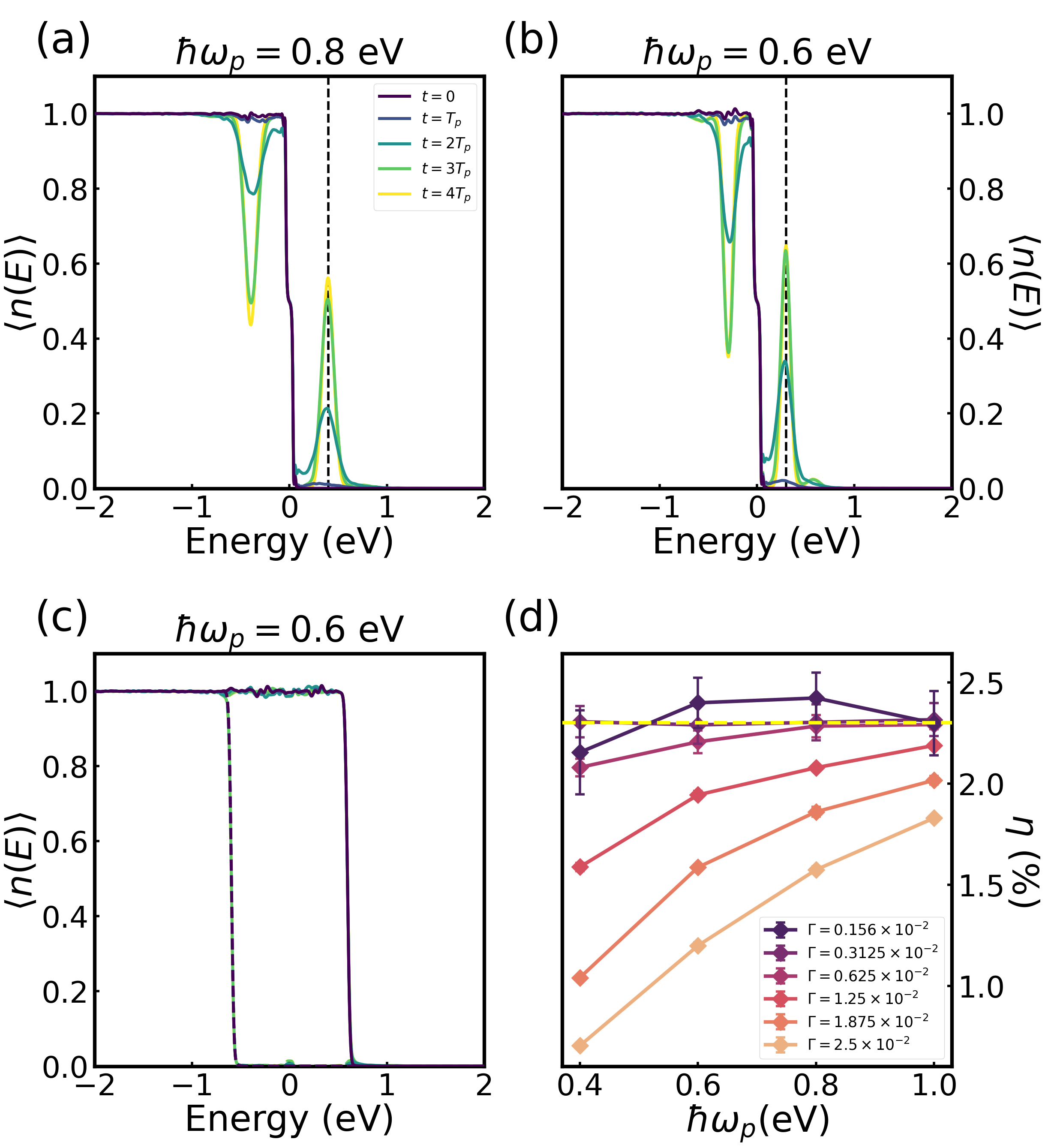}
\caption{Time evolution of the graphene carrier occupation for photon energies (a) $\hbar\omega_p = 0.8$ eV and (b) $0.6$ eV, with $\mu_0 = 0$. The dashed lines indicate the expected absorption peak at $\hbar\omega_p/2$. (c) Evolution of $n$- and $p$-doped distributions with $\mu_0 = \pm 0.6$ eV (solid and dashed lines, respectively), indicating Pauli blocking. (d) Absorption efficiency vs.\ photon energy for different optical intensities, with the universal optical absorption of monolayer graphene $\eta\approx2.3\%$ highlighted by the yellow dashed line. In all cases, $T_0  = 0$ K with a pulse envelope
$P(t)=\sech\left[ (t-2T_p) / \gamma T_p \right]$,
where $T_p = 2\pi / \omega_p$ and $\gamma \approx 0.5673$.}
\label{fig:Fig1}
\end{figure}

Equations \eqref{eq:occupation} and \eqref{eq:evolution2} require evaluation of the functions $\delta(\ham(t)-\eps)$ and $\exp(-\x{i} \ham(t) \dt / \hbar)$. We efficiently evaluate these by expanding them as a series of Chebyshev polynomials \cite{WeisseKPM, roche2005quantum, Fan2021linear,Ishii2010}, and the method thus boils down to a series of multiplies between the matrix $\ham(t)$ and the vectors $\ket{\psi_r(t)}$ and $\ket{\psi_r^{\F}(t)}$ (see SM for details about the Chebyshev expansion \cite{SupplementaryMaterial}). For sparse tight-binding Hamiltonians, the total simulation cost is then $\mathcal{O}(R M N_t N)$, where $M$ is the number of Chebyshev polynomials and $N_t$ is the number of time steps. $R$, $M$, and $N_t$ are decoupled from $N$, and the method thus scales linearly with the number of sites \cite{Fan2021linear}, allowing for the calculation of time-resolved nonequilibrium dynamics in systems reaching the experimental scale. Below, we simulate samples with $512 \times 512 \times 4 \approx 1$ million and $64 \times 128 \times 80 \approx 655\,000$ carbon atoms for graphene and NPG, respectively, and for Fig.\ \ref{fig:Fig1}(d) we used $2048 \times 2048 \times 4 \approx 16$ million carbon atoms. All of our samples have periodic boundaries, and are treated within the single-particle picture. 

Here we point out that the major difference between this approach and prior linear-scaling techniques, as outlined in Fan et al.\ \cite{Fan2021linear}, is in the treatment of energy transfer and out-of-equilibrium dynamics. Prior methods focused on the evolution of closed systems where the energy distribution of the carriers does not change over time. In contrast, here we explicitly account for the absorption or emission of energy and its impact on the time-dependent carrier distribution. Such an approach is necessary to handle the effect of external time-dependent interactions such as light, heat, mechanical driving, etc., enabling the study of nontrivial out-of-equilibrium quantum dynamics in large-scale disordered systems.

\textit{Optical excitations in graphene} -- To demonstrate the methodology, we examine the evolution of the carrier distribution in graphene when illuminated by a linearly-polarized optical pulse. We consider a minimal tight-binding model of graphene \cite{FoaTorres_Roche_Charlier_2014}, 
$\ham = - \sum_{\langle i,j\rangle}t_{ij} c^{\dagger}_{i} c_{j} + \sum_{i} V_i c^{\dagger}_i c_i,$
where the first term is the nearest-neighbor hopping, with $t_{ij} = t = 2.71$ eV. The second term represents electrostatic disorder. Here we use Anderson disorder, where $V_i$ is a random onsite potential evenly distributed in $\left[-W/2,W/2\right]$ with $W$ the disorder strength.

To include the linearly-polarized optical field, we define a vector potential $\bm{A}(t)=A_0P(t)\sin(\omega_p t)\hat{y}$, where $A_0$ is the amplitude, $P(t)$ is the envelope of the optical pulse, and $\hbar\omega_p$ is the photon energy \cite{dal2017one}. Using the Peierls substitution, the nearest-neighbor hopping becomes
\begin{equation}
t_{ij}(t) = t \exp{\left( i\frac{2\pi}{\Phi_0}\int_{\bm{r}_i}^{\bm{r}_j} d\bm{r} \cdot \left[A_0P(t)\sin(\omega t)\hat{y}\right]\right),}\label{eq:tdephop}
\end{equation}
where $\bm{r}_i$ is the position of carbon site $i$ and $\Phi_0=h/e$ is the magnetic flux quantum. Below we set $A_0=\Gamma\Phi_0  /2a$, where $\Gamma$ is a free parameter to regulate the field intensity and $a = 2.46$ \angstrom is the graphene lattice constant.

\begin{figure*}[th]
\centering
\includegraphics[width=0.9\linewidth]{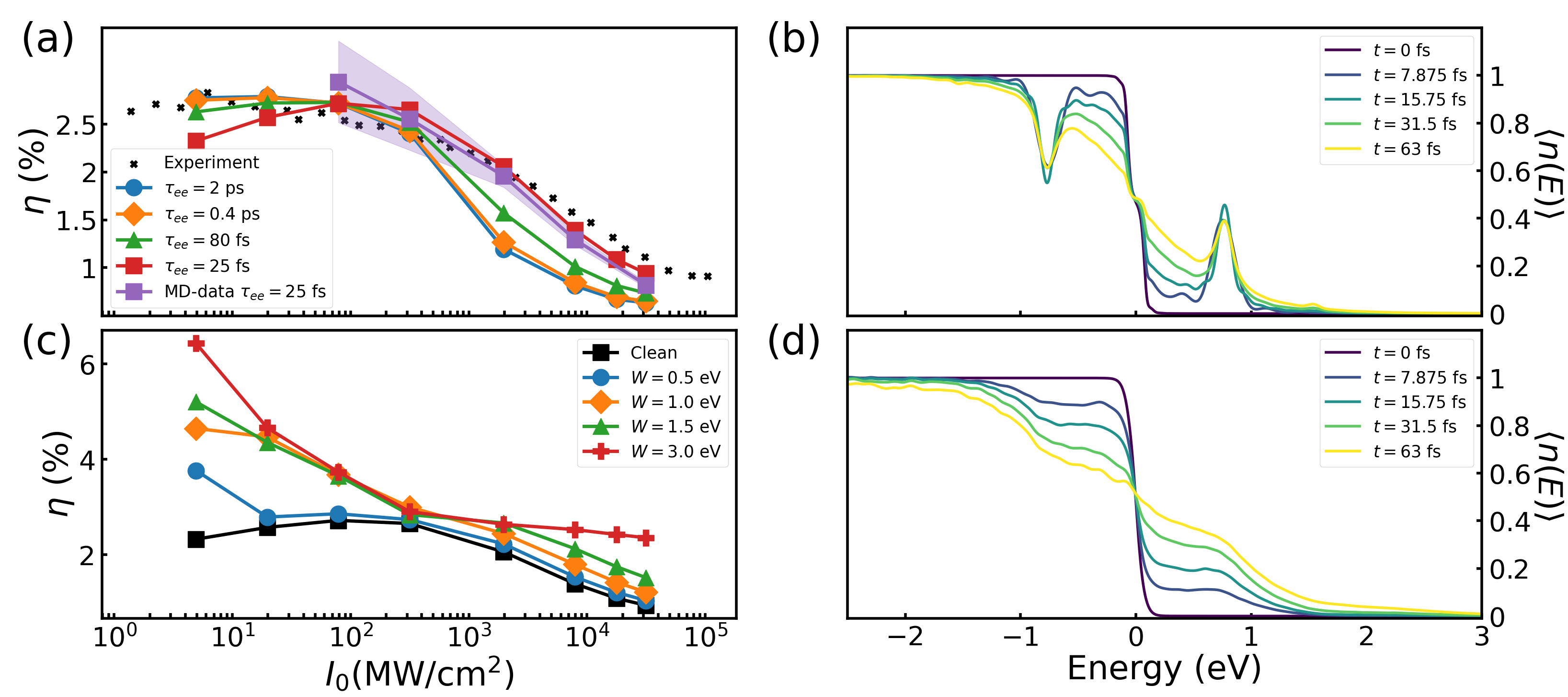}
\caption{(a) Energy absorption efficiency vs.\ light intensity for clean graphene with $\mu_0=-5$ meV and $T_0 = 300$ K, under a 63-fs optical pulse with $\hbar\omega_p = 1.55$ eV. Black symbols are experimental data \cite{baek2012efficient}, and colored lines are simulations with varying $\tee$. The solid line and the shaded area for the MD data correspond, respectively, to the mean optical absorption and its standard deviation over $10$ graphene samples. (b) Time-dependent carrier distribution at high intensity ($I_0 =  3.02\times10^4$ MW/cm$^2$) for $\tee = 25$ fs. (c) Energy absorption efficiency vs.\ light intensity in the presence of Anderson disorder, for the same conditions as in panel (a), with $\tee = 25$ fs. (d) Time-dependent carrier distribution at high intensity ($I_0 =  3.02\times10^4$ MW/cm$^2$) for Anderson disorder strength $W=3$ eV.}
\label{fig:Fig2}
\end{figure*}

To validate the methodology, we first consider clean graphene without relaxation, $W = 0$ and $\tau \rightarrow \infty$. We apply an optical pulse with an envelope
$P(t)=\sech\left[ (t-2T_p) / \gamma T_p \right]$,
where $T_p = 2\pi/\omega_p$ is the period of field oscillation and $\gamma \approx 0.5673$, such that the full width of $P^2(t)$ at half maximum is equal to $T_p$ \cite{keller2021ultrafast}. We consider a simulation time of $t = 0 \rightarrow 4T_p$.

Figures \ref{fig:Fig1}(a) and (b) show the time evolution of an initially undoped carrier distribution at zero temperature, when illuminated by pulses with $\hbar\omega_p = 0.8$ eV and $0.6$ eV, respectively. In both cases the field amplitude was set to $\Gamma = 2.5\times10^{-2}$. Absorption peaks develop at $\pm\hbar\omega_p/2$, as predicted by theory \cite{xing2010physics}, with their width attributed to the finite pulse duration and the numerical broadening inherent in the Chebyshev expansion of $\delta(\ham(t)-\eps)$ \cite{WeisseKPM, Fan2021linear}. It is also worth noting that the height of the absorption peak increases with decreasing photon energy; because of its longer duration, the total energy delivered by the $0.6$-eV pulse is greater than for $0.8$ eV.

Figure \ref{fig:Fig1}(c) confirms that the method obeys the Pauli exclusion principle. Here we show the evolution of $n$- and $p$-doped carrier distributions, with $\mu_0 = \pm0.6$ eV, under an optical pulse with $\hbar\omega_p = 0.6$ eV. This figure shows that for $|\mu_0| > \hbar\omega_p/2$, Pauli blocking forbids any carrier excitation from the valence to the conduction band.

Finally, we show that the method recovers the universal optical absorption of graphene \cite{nair2008fine, OpticalCondGraphene}. To quantify this, we calculate the energy absorption efficiency,
\begin{equation}
\eta(t) = \frac{E_\x{el}(t) - E_\x{el}(0)}{E_\x{opt}(t)},
\end{equation}
where $E_\x{el}(t) = \braket{\psi_r(t) | \ham(t) | \psi_r^{\F}(t)}$ is the energy of the electrons and $E_\x{opt}(t) = A_\mathrm{S} \epsilon_0 c \int_0^t |\partial_t\bm{A}(t)|^2 dt$ is the total energy irradiated over the sample at time $t$, where $A_\mathrm{S}$, $\epsilon_0$ and $c$ are the sample area, vacuum permittivity, and speed of light, respectively \cite{strattonEMtheory}. In Fig.\ \ref{fig:Fig1}(d) we plot $\eta(t = 4T_p)$ as a function of $\hbar\omega_p$ for different optical intensities $\Gamma$. Note that because of the form of the envelope $P(t)$, lower $\hbar\omega_p$ corresponds to a longer pulse and higher total optical energy delivered to the graphene. In the limit of low total optical energy (smaller $\Gamma$ and higher $\hbar\omega_p$), the absorption efficiency reaches the predicted universal value of $\sim$$2.3\%$, indicated by the horizontal dashed line. Meanwhile, at higher optical energy delivered to the graphene the efficiency decreases due to the onset of saturable absorption arising from Pauli blocking.

\textit{Saturable absorption in graphene} -- Having verified the basic features of optical absorption, we now use the method to compare to measurements of saturable absorption in graphene \cite{baek2012efficient}. To mimic the experimental conditions we let $\mu_0 = -5$ meV and $T_0 = 300$ K, we apply a 63-fs optical pulse with energy $\hbar\omega_p = 1.55$ eV, and we add an absorption background of $0.4\%$ to our results. We also include electron-electron scattering with thermalization time $\tee$, see Eqs.\ \eqref{eq:evolution} and \eqref{eq:evolution2}.

In Fig.\ \ref{fig:Fig2}(a) we plot the energy absorption efficiency as a function of optical intensity. The black symbols are the experimental results, extracted from Fig.\ 3 of Ref.\ \citenum{baek2012efficient} by converting transmission to absorption and fluence to power density, and the colored lines are our simulations. Each line corresponds to a different value of $\tee$. In agreement with the measurements, higher optical intensity leads to a reduction of absorption efficiency, owing to optical bleaching. Meanwhile, reducing $\tee$ improves absorption efficiency at higher optical intensity, as faster thermalization redistributes the absorption peaks and suppresses Pauli blocking \cite{marini2017theory}. We find the best agreement between our simulations and the experiments when $\tee = 25$ fs, which falls within the estimates for carrier-carrier relaxation times in graphene \cite{brida2013ultrafast}.

To analyze the impact of thermally generated hopping noise, we used $10$ different $80 \times 80$ nm${}^2$ graphene samples generated with molecular dynamics simulations at $300$ K, as implemented in the LAMMPS package \cite{thompson2022lammps}, and replicated them in a $2 \times 2$ superlattice to match the number of atoms used in the flat system.
We also included a small Anderson disorder ($W=1$ $\upmu$eV) to break the periodicity. Comparing the average saturable absorption with that of flat graphene, we find that thermal fluctuations at 300 K appear to have no impact on saturable absorption in otherwise disorder-free graphene.

Figure \ref{fig:Fig2}(b) shows the evolution of the carrier distribution at high optical intensity for $\tee = 25$ fs. There is a sizable buildup of carriers at $\pm\hbar\omega/2$, resulting in Pauli blocking and a reduction of the absorption efficiency. One can also see the impact of a short $\tee$, with the formation of a thermal distribution for $|\epsilon|<\hbar \omega_p/2$. However, even with this short $\tee$, at high intensities carrier-carrier scattering cannot fully relax the absorption peaks, resulting in bleaching and poor absorber performance.

\begin{figure}[ht]
\centering
\includegraphics[width=\linewidth]{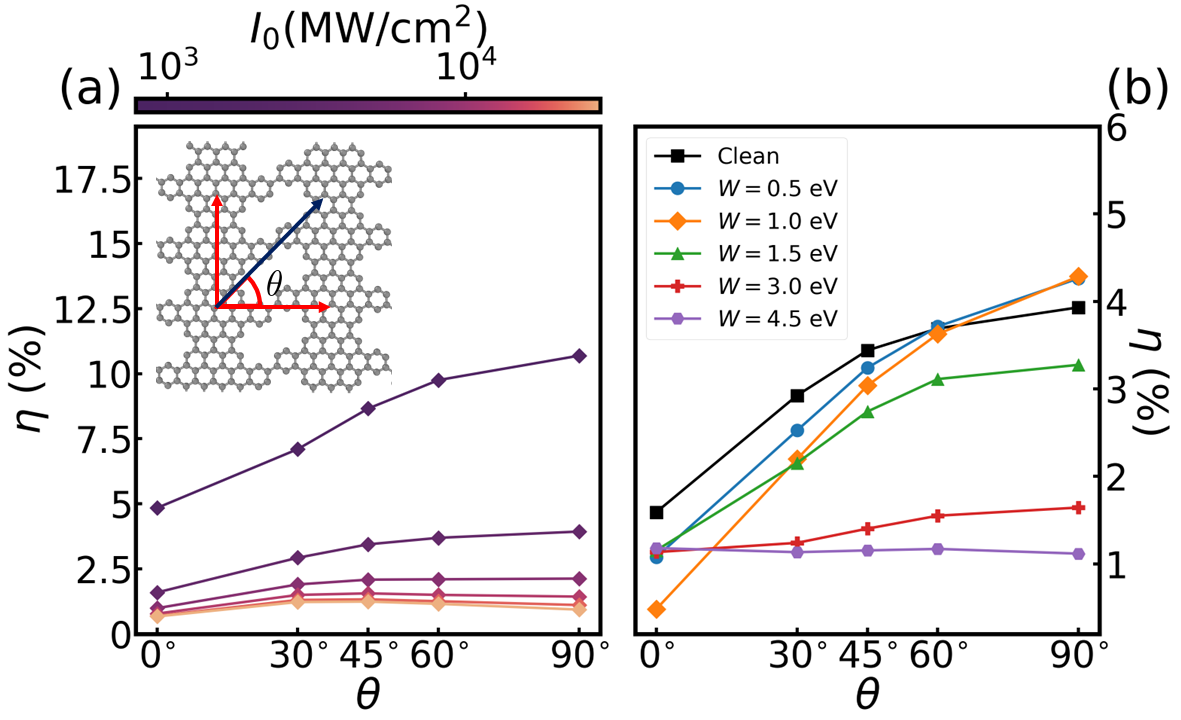}
\caption{(a) Anisotropic energy absorption of NPG for $\mu_0 = -0.1$ eV, $T_0 = 300$ K and $\tee = 25$ fs under a $97.65$-fs pulse with $\hbar\omega_p = 1$ eV at various optical intensities. Inset: Schematic of NPG used in our simulations. The red arrows are the ribbon axes and the blue arrow is the direction of light polarization. (b) Angle-dependent energy absorption for disordered NPG under the same conditions, for $I_0 = 3.142 \times 10^3$ MW/cm$^2$ and various Anderson disorder strengths $W$.}
\label{fig:Fig3}
\end{figure}

Next, we consider the same situation in the presence of real-space disorder. Figure \ref{fig:Fig2}(c) shows the absorption efficiency as a function of light intensity for various Anderson disorder strengths. Remarkably, absorption efficiency is enhanced with increasing disorder strength. This result correlates with the broadening of the density of states, which renormalizes the Fermi velocity and increases the number of states available for carrier excitation \cite{SupplementaryMaterial}. Such disorder-induced absorption enhancement has also been seen in photocurrent and photoluminescence measurements, as well as many-body theoretical calculations, of graphene/MoS${}_2$ heterobilayers containing localized defects \cite{hotger2023photovoltage}. In these systems, sulfur vacancies in the MoS${}_2$ enhance the optical absorption in graphene by creating additional absorption pathways enabled by the subgap localized disorder states.
Meanwhile, even though absorption increases with disorder, we still see bleaching and a reduction of absorption with increasing optical intensity.

Figure \ref{fig:Fig2}(d) shows the evolution of the carrier distribution at high optical intensity, for $W = 3$ eV and $\tee = 25$ fs. Compared to panel (b), one can see that the carrier dynamics of the disordered system drastically differ from the clean one. Specifically, disorder leads to a much more effective thermalization of carriers, as the broadening opens optical absorption pathways other than $-\hbar\omega_p/2 \rightarrow +\hbar\omega_p/2$ \cite{song2015energy}. This enables enhanced absorption at low optical intensities while also suppressing bleaching at high intensities, and is similar to a ``pre-thermalization'' effect examined in Ref.\ \citenum{yang2017reassessing}.

\textit{Anisotropic absorption in nanoporous graphene} -- We now apply the method to NPG, to examine the impact of saturable absorption and disorder on its optical anisotropy. The inset of Figure \ref{fig:Fig3}(a) shows the structure of NPG. We model its electronic properties with the graphene Hamiltonian, which has been shown to reproduce the band structure obtained from DFT \cite{calogero2018electron, calogero2019quantum, alcon2024tailoring}.
To analyze optical anisotropy, we define a vector potential $\bm{A}(t) = A_0 \sin(\omega t) \left[ \cos(\theta)\hat{x} + \sin(\theta)\hat{y} \right]$, where $\theta$ is the direction of light polarization, see Fig.\ \ref{fig:Fig3}(a). In our simulations, we let $\mu_0 = -0.1$ eV, $T_0 = 300$ K, $\tee = 25$ fs, and we apply a $97.65$-fs pulse with energy $\hbar\omega_p = 1$ eV.

Figure \ref{fig:Fig3}(a) shows the angle-dependent absorption in clean NPG at different optical intensities. Anisotropy is clearly visible at low intensity, with the ratio $\eta(90^\circ) / \eta(0^\circ) \approx 2.2$. Meanwhile, $\eta(0^\circ) > 0$ in all cases, consistent with a small but finite charge transport perpendicular to the ribbons \cite{calogero2018electron, calogero2019quantum, alcon2024tailoring}.
On the other hand, increasing the light intensity triggers optical bleaching and suppresses the anisotropy, down to a ratio of $\sim$$1.4$ at the highest intensity. Interestingly, at this intensity, we find that energy absorption is maximized at $\theta \approx 45^{\circ}$.

To analyze the impact of disorder, we fix the light intensity to $I_0 = 3.142 \times 10^3$ MW/cm$^2$ and calculate the angle-dependent absorption for various Anderson strengths. As seen in Fig.\ \ref{fig:Fig3}(b), weak disorder significantly enhances optical anisotropy, up to a ratio of $\sim$$9$ for $W = 1$ eV. Most of this enhancement comes from a strong suppression of absorption perpendicular to the ribbons. This correlates with the rapid onset of electron localization in this direction, previously observed in quantum transport simulations \cite{alcon2024tailoring}.

In contrast, weak disorder enhances optical absorption parallel to the ribbons, similar to graphene. However, for $W \geq 1.5$ eV, both absorption and anisotropy are then suppressed, indicating the onset of electron localization. For highly disordered NPG ($W=4.5$ eV), absorption is reduced to $\sim$$1\%$ and becomes fully isotropic, indicating a subtle interplay between disorder and optical anisotropy that may be exploitable for sensing applications.

\textit{Summary and conclusions} -- We have developed a linear-scaling numerical framework that enables the study of nonequilibrium carrier dynamics in systems reaching experimental length scales, with a fully nonperturbative treatment of defects and disorder. Thanks to its scalability, this methodology can be naturally integrated with AI-assisted Hamiltonian-generation schemes trained on accurate first-principles calculations, offering a practical route to describe far-from-equilibrium phenomena in large-scale systems with realistic disorder, as commonly encountered in experiments. Additionally, its compact formulation can be extended to compute the time-dependent evolution of a broad range of observables under driving fields with arbitrary temporal profiles, without additional computational cost. We have validated the approach by capturing both universal absorption and saturable absorption in graphene, and then used it to predict disorder-induced enhancement of energy absorption and optical anisotropy in graphene and nanoporous graphene (NPG). Beyond the optical response of carbon-based two-dimensional materials, the method is directly generalizable to any time-dependent tight-binding Hamiltonian, making it broadly applicable to the study of far-from-equilibrium electron dynamics in a wide variety of materials and excitation conditions.

\begin{acknowledgments}
L.M.C.\ acknowledges funding from the Ministerio de Ciencia e Innovación de Espa\~na under grant no.\ PID2019-106684GB-I00 / AEI / 10.13039/501100011033, FJC2021-047300-I, financed by MCIN / AEI / 10.13039/501100011033 and the European Union ``NextGenerationEU/PRTR''. L.M.C.\ and A.W.C.\ acknowledge funding from the U.S.\ Army Research Office under grant no.\ W911NF-21-1-0004. We acknowledge funding from Ministerio de Ciencia e Innovación under grant no.\ PID2019-106684GB-I00 financed by MCIN / AEI / 10.13039/501100011033 and Grant PID2022-138283NB-I00 funded by MICIU/AEI/ 10.13039/501100011033 and by “ERDF/EU”. ICN2 is funded by the CERCA Programme/Generalitat de Catalunya and supported by the Severo Ochoa Centres of Excellence programme, Grant CEX2021-001214-S, funded by MCIN / AEI / 10.13039.501100011033.
\end{acknowledgments}

\onecolumngrid
\
\renewcommand{\theequation}{S\arabic{equation}}
\renewcommand{\thefigure}{S\arabic{figure}}

\newpage
{\center \large \bf Supplementary Material for \\ ``Real-Time Out-of-Equilibrium Quantum Dynamics in Disordered Materials''}

\section{Expansion of spectral functions using Chebyshev polynomials}

The nonequilibrium dynamics in the main text were calculated using the kernel polynomial method (KPM) \cite{silver1996kernel, WeisseKPM, Fan2021linear}. This procedure consists of expanding functions of the Hamiltonian as a series of powers of the Hamiltonian, specifically as a series of Chebyshev polynomials. This avoids the need to diagonalize the Hamiltonian, which becomes computationally prohibitive for even moderate system sizes.
In what follows, we present the general method and demonstrate its application in approximating the density of states $\delta(\ham-\varepsilon)$ \cite{silver1996kernel}, the Fermi operator $\F(\ham,T,\mu) = [ 1 + \exp((\ham-\mu) / k_\mathrm{B}T) ]^{-1}$ \cite{baer1997sparsity, baer1997chebyshev, ordejon1998order, goedecker1999linear}, and the time evolution operator $\hat{U}(\Delta t)=\exp(-\mathrm{i} \ham \Delta t / \hbar)$ \cite{wang1998time, fehske2009numerical}.

First we rescale the Hamiltonian $\ham$ and the energies $\varepsilon$ to the interval $(-1,1)$, where the Chebyshev polynomials are bounded, to guarantee the convergence of the series expansion. The rescaled quantities are $\tH = (\ham-b)/a$ and $\tep = (\varepsilon-b)/a$. Here $a = (\varepsilon_\mathrm{max} - \varepsilon_\mathrm{min})/2$ and $b = (\varepsilon_\mathrm{max} + \varepsilon_\mathrm{min})/2$, where $\varepsilon_\mathrm{max}$ and $\varepsilon_\mathrm{min}$ represent the maximum and minimum energies of the spectrum, respectively.

As shown in Refs.\ \onlinecite{WeisseKPM, Fan2021linear}, an arbitrary function of the form $f(\tep,\tH)$ can then be written as
\begin{equation}
f(\tep,\tH) = \sum\limits_{m=0}^{\infty} \Gamma_m(\tep) T_m(\tH),
\label{eqn:chebexpspectral}
\end{equation}

\noindent where
\begin{equation}
\Gamma_m(\tep) = (2-\delta_{m0}) \int\limits_{-1}^{1} \mathrm{d}E \frac{f(\tep,E) T_m(E)}{\pi \sqrt{1-E^2}},
\label{eqn:chebexpcoef}
\end{equation}

\noindent and $T_m(x) = \cos(m\arccos(x))$ are the Chebyshev polynomials of the first kind. These can be efficiently computed from the recurrence relation $T_{m+1}(x) = 2x T_m(x) - T_{m-1}(x)$, with $T_0 = 1$ and $T_1 = x$. In Eq.\ \eqref{eqn:chebexpspectral}, all the information associated with the energy dependence of $f(\tep,\tH)$ has thus been separated from the information of the Hamiltonian.

In practical applications, it is always necessary to truncate the series at some finite order $M$. The truncation of the series leads to reduced precision and Gibbs oscillations, especially when the function is not continuously differentiable. However, one can dampen these oscillations and recover the precision of the expansion by modifying the moments $\Gamma_m$ with a kernel $g_m$, $\Gamma_m \rightarrow g_m \Gamma_m$. In the manuscript, we used the Jackson kernel given by 
\begin{equation}
g_m = \frac{1}{M+1} \left[ (M-m+1) \cos\left(\frac{\pi m}{M+1}\right) + \sin\left(\frac{\pi m}{M+1}\right)\cot\left(\frac{\pi}{M+1}\right) \right].
\end{equation}

\noindent This kernel has been recognized as optimal for most situations since it is positive and normalized, preserving the integral of the expanded function \cite{WeisseKPM}.

The Chebyshev polynomial expansion of the density of states is then obtained from the substitution $f(\tep,E) = \delta(E-\tep)$ in Eq.\ \eqref{eqn:chebexpcoef},
\begin{equation}
\delta(\tH-\tep) \approx \frac{1}{\pi\sqrt{1-{\tep}^2}} \sum\limits_{m=0}^M (2-\delta_{m0}) g_m T_m(\tep) T_m(\tH). \label{eqn:DOSexpansion}
\end{equation}

\noindent A similar procedure can be followed for the Fermi operator, $\F(\ham,T,\mu)$. However, for finite temperature, the coefficients $\Gamma_m(\tep)$ must determined through numerical integration.

Since the Chebyshev polynomial expansion is efficient when the function is infinitely differentiable, its application to the time evolution operator does not require a damping kernel. Thus, the polynomial expansion of the time evolution operator reads as
\begin{equation}
\hat{U}(\Delta t) = \exp\left( -\mathrm{i} \frac{\ham}{\hbar} \Delta t \right) = \exp\left( -\mathrm{i} \frac{b}{\hbar} \Delta t \right)\exp\left( -\mathrm{i} \frac{a \tH}{\hbar} \Delta t \right) \approx \exp\left( -\mathrm{i} \frac{b}{\hbar} \Delta t \right) \sum\limits_{m=0}^M U_m \left( \frac{a}{\hbar} \Delta t \right) T_m(\tH),
\end{equation}
\noindent where $U_m(x) = (2-\delta_{m0})(-\mathrm{i})^m J_m(x)$ with $J_m(x)$ $m$th order Bessel function of first kind. In practice, the number of expansion moments need not be the same for $\hat{U}$ as for $\delta$ or $\F$. For further details on the  Chebyshev expansion techniques and their application to quantum transport, we refer the readers to Ref.\ \onlinecite{Fan2021linear}.

\section{Equivalence with Liouville Equation and Carrier Conservation}

As detailed in the main text, we proposed a new method for studying the dynamics of excited carriers and their transition to equilibrium in large-area disordered systems. During the time evolution we track two state vectors. The first, $|\psi_r(t)\rangle$, is a random phase vector that contains all the information related to the available energy states and evolves under the Schr\"odinger equation. The second, $|\psi^{\F}_r(t)\rangle$, carries the history of the initial electronic distribution and evolves under the modified Eq.\ (2) of the main text. In the following, we will show that this modified equation lead us to an expression similar to the relaxation time approximation for the trace of the Liouville equation.

From the main text, the time evolutions of $|\psi_r(t)\rangle$ and $|\psi^{\F}_r(t)\rangle$ are given by
\begin{align}
    \partial_{t} \ket{\psi_r(t)} &= -\mathrm{i} \frac{\ham(t)}{\hbar} \ket{\psi_r(t)} \nonumber \\
    \partial_{t} \ket{\psi^{\F}_r(t)} &= -\mathrm{i} \frac{\ham(t)}{\hbar} \ket{\psi^{\F}_r(t)} - \frac{1}{\tau}\left( \ket{\psi^{\F}_r(t)} - \Neq(t)\ket{\psi_r(t)} \right).
    \label{eq:time-ev1}
\end{align}

\noindent Here, $\Neq(t)$ represents the instantaneous equilibrium distribution toward which the electronic system will relax, on a time scale given by $\tau$. Next, we can write the time evolution of the matrix $|\psi^{\F}_r\rangle \langle\psi_r(t)|$ as
\begin{align}
    \partial_{t} \left( \ket{\psi^{\F}_r(t)} \bra{\psi_r(t)} \right) = \partial_{t} \ket{\psi^{\F}_r(t)} \bra{\psi_r(t)} + \ket{\psi^{\F}_r(t)} \partial_t \bra{\psi_r(t)}.
    \label{eq:time-ev2}
\end{align}

\noindent Substituting Eq.\ \eqref{eq:time-ev1} into \eqref{eq:time-ev2} and using the properties of the trace, we find that the time evolution of the trace of $\ket{\psi^{\F}_r(t)} \bra{\psi_r(t)}$ is given by
\begin{align}
\partial_t \trace{\ket{\psi^{\F}_r(t)} \bra{\psi_r(t)}} = -\frac{1}{\tau} \left( \trace{\ket{\psi^{\F}_r(t)} \bra{\psi_r(t)}} - \trace{\Neq(t) \ket{\psi_r(t)} \bra{\psi_r(t)}} \right)
\end{align}

\noindent Following Refs.\ \onlinecite{ordejon1998order, goedecker1999linear}, we identify that $\ket{\psi^{\F}_r(t)} \bra{\psi_r(t)} = \rho(t)$ and $\Neq(t) \ket{\psi_r(t)} \bra{\psi_r(t)} = \rho_\mathrm{eq}(t)$. Consequently, the time evolution of the electronic density matrix in our system reads as
\begin{align}
    \partial_{t}\trace{\rho(t)} = -\frac{1}{\tau} \trace{\rho(t) - \rho_\mathrm{eq}(t)}.
\end{align}

This equation shows the connection of our method with the Liouville equation, capturing the transition toward a new equilibrium state with arbitrary energy and carrier concentration.

\section{Determination of the equilibrium distribution}

As explained above and in the main text, the carrier density transitions toward the instantaneous equilibrium distribution modeled by the operator $\Neq(t)$, whose form is governed by the physics of the system. In this work we focused on optical absorption in graphene and nanoporous graphene, on time scales where carrier-carrier scattering is the dominant relaxation mechanism. Carrier-carrier scattering redistributes the carrier distribution toward a thermal (Fermi-Dirac) distribution, and thus we let $\Neq(t) = \F \left( \ham(t) , T(t) , \mu(t) \right)$.

At each time step, the temperature $T(t)$ and chemical potential $\mu(t)$ are chosen to ensure conservation of energy and carrier density. The former is accomplished by ensuring the total energy of $\Neq(t)$ equals the total energy of the instantaneous carrier distribution,
\begin{equation}
\braket{\psi_{r}(t) | \ham(t) \Neq(t) | \psi_r(t)} = \braket{\psi_{r}(t) | \ham(t) | \psi^{\F}_r(t)}.
\label{eq:energyconsev}
\end{equation}

\noindent Simultaneously, we ensure carrier conservation by imposing the constraint $n_\mathrm{h}(t) - n_\mathrm{e}(t) = n_\mathrm{h}(0) - n_\mathrm{e}(0)$, where $n_\mathrm{h}$ and $n_\mathrm{e}$ are number of holes and electrons, respectively,
\begin{align}
    n_\mathrm{h}(t) &= \int\limits_{-\infty}^{\varepsilon_\mathrm{cnp}} \braket{\psi_r(t) | \delta(\ham(t)-\varepsilon) \left[ 1-\F\left(\ham(t),T(t),\mu(t)\right) \right] | \psi_r(t)} \mathrm{d}\varepsilon, \\
    n_\mathrm{e}(t) &= \int\limits_{\varepsilon_\mathrm{cnp}}^{\infty} \braket{\psi_r(t) | \delta(\ham(t)-\varepsilon) \F\left(\ham(t),T(t),\mu(t)\right) | \psi_r(t)} \mathrm{d}\varepsilon,
\end{align}
and $\varepsilon_\mathrm{cnp}$ is the energy at which $n_\mathrm{h}(t) = n_\mathrm{e}(t)$.

\section{Density of States of Disordered Graphene and Nanoporous Graphene}

\begin{figure}[h]
\includegraphics[width=0.8\linewidth]{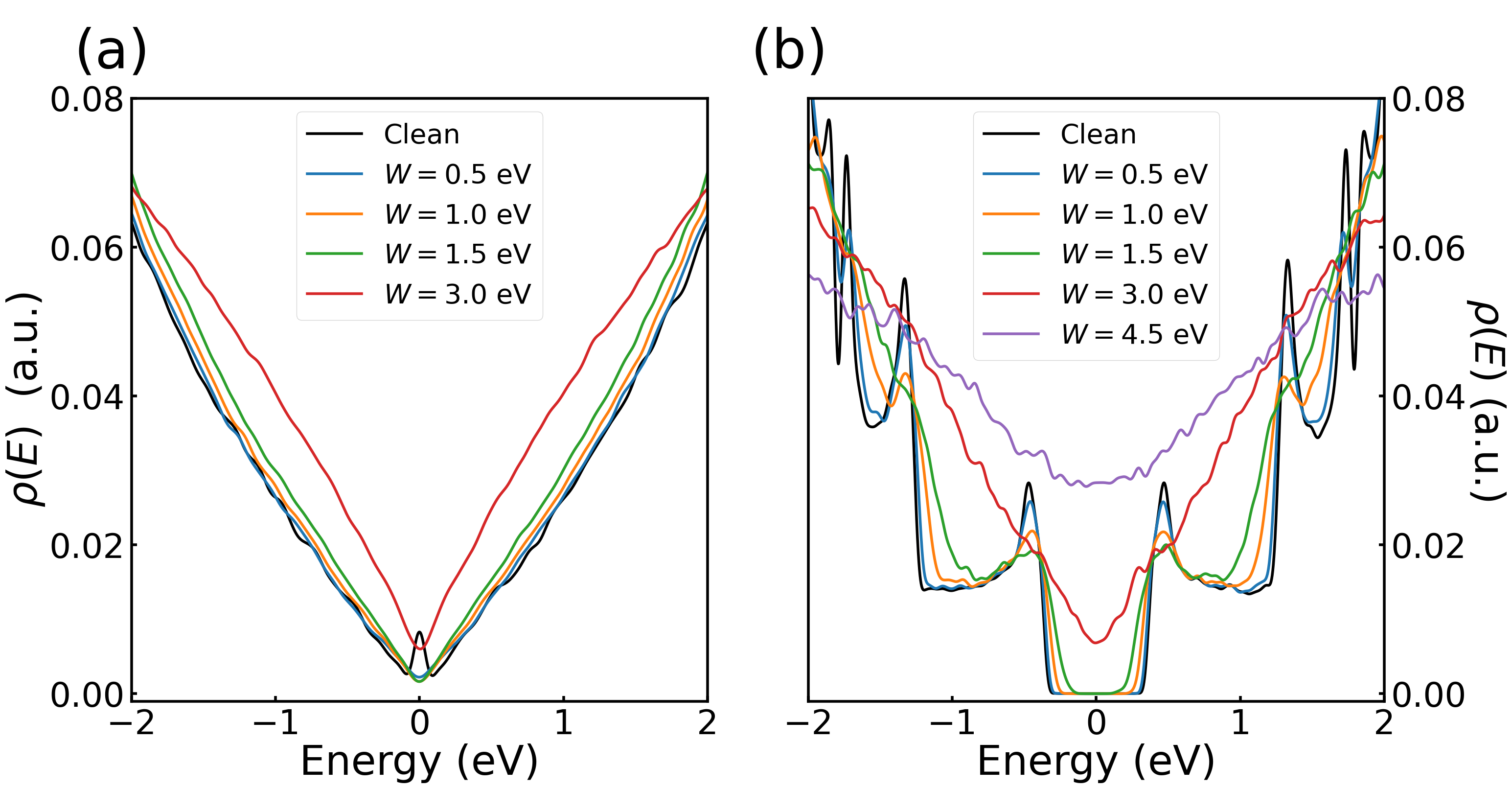}
\caption{(a) Density of states of graphene with $512\times512\times4$ carbon atoms for various Anderson disorder strengths $W$, computed with $M=1024$ Chebyshev polynomials. (b) Density of states of a nanoporous graphene system composed of $64\times128\times80$, using $M=1536$ Chebyshev polynomials.}
\label{fig:DOSSupmat}
\end{figure}

As discussed in the main text for both graphene and nanoporous graphene (NPG), the presence of disorder changes the number of states available for light-induced transitions, and thus their energy absorption. Figure \ref{fig:DOSSupmat}(a) shows the density of states of the graphene samples we used in our simulations, consisting of $512\times512\times4$ carbon atoms with open boundary conditions. For the clean case there is a peak in the density of states at $E=0$ due to edge states. However, this peak is washed away as the Anderson disorder strength ($W$) increases. Moreover, as discussed in the main text, due to the action of the disorder one can observe that the number of states is increased in the low-energy region where the optical transitions occur, particularly for $W=3$ eV. The increased number of states due to disorder correlates with the enhanced optical absorption we found and with the distinct carrier dynamics at high intensities.

In Figure \ref{fig:DOSSupmat}(b), we show the density of states of the NPG structure we used in our calculations, consisting of $64\times128\times80$ carbon atoms with periodic boundary conditions. In Fig.\ 3 of the main text, we found that the absorption anisotropy increased for disorder strengths up to $W=1$ eV. Observing the changes in the DOS, it is clear that the main features of NPG remain at small disorder, albeit with a small broadening. In contrast, at larger disorder strengths the energy gap becomes populated, and the quasi-1D features of the density of states at higher energies are completely broadened, which correlates with the loss of optical absorption anisotropy.

\end{document}